\documentclass[runningheads]{llncs}
\usepackage[T1]{fontenc}
\usepackage{algpseudocode}
\usepackage{amsmath}
\usepackage[normalem]{ulem}
\usepackage{float}
\usepackage{amssymb}
\usepackage{graphicx}
\usepackage{subcaption}
\usepackage{tabularx}
\usepackage{multirow}
\usepackage{diagbox}
\usepackage{bm} 
\usepackage{hyperref}
\usepackage{academicons}
\usepackage{orcidlink}
\usepackage{caption} 
\usepackage[table]{xcolor}
\usepackage{sidecap}
\usepackage{makecell}
\usepackage[most]{tcolorbox}
\usepackage{array}
\usepackage{pifont}
\usepackage{wasysym}
\usepackage{rotating}
\usepackage{adjustbox}
\usepackage{wrapfig} 
\definecolor{neongreen}{RGB}{0,255,0}
\begin{document}
\title{Residual Risk Assessment in Benign Code: How Far Are We? A Multi-Model Semantic and Structural Similarity Approach}
\titlerunning{Residual Risk Assessment in Benign Code: How Far Are We?}
%
\author{Mohammad Farhad\inst{1,2}\orcidID{0009-0002-1788-6082} \and
Shuvalaxmi Dass\inst{1,2}\orcidID{0000-0001-9254-8134}}
\authorrunning{M. Farhad et al.}
%
\institute{University of Louisiana at Lafayette, Lafayette LA 70504, USA \and
AICSIL Research Lab\\
\email{\{mohammad.farhad1,shuvalaxmi.dass\}@louisiana.edu}}

\newcommand{\greenCircle}{\tikz\fill[green] (0,0) circle (0.6ex);}
\newcommand{\redCircle}{\tikz\fill[red] (0,0) circle (0.6ex);}
\newcommand{\yellowCircle}{\tikz\fill[yellow] (0,0) circle (0.6ex);}

\newcommand{\outlier}{\textcolor{black}{\large $\bullet$}}
\newcommand{\normalpt}{\textcolor{black}{\large $\circ$}}

\newcommand{\lightredCircle}{\tikz\fill[red!40] (0,0) circle (0.6ex);}
\newcommand{\lightyellowCircle}{\tikz\fill[yellow!50] (0,0) circle (0.6ex);}
\newcommand{\lightgreenCircle}{\tikz\fill[green!40] (0,0) circle (0.6ex);}
\maketitle              

\begin{abstract}
Software security assurance relies on effective vulnerability detection and patching, yet determining whether a patch fully eliminates risk remains an underexplored challenge. Existing vulnerability benchmarks often treat patched functions as inherently benign, overlooking the possibility of residual security risks. In this work, we analyze vulnerable–benign C/C++ function pairs from PrimeVul, a benchmark dataset of real-world vulnerabilities. We use multiple code language models (Code LMs) to capture semantic similarity, complemented by Tree-sitter–based abstract syntax tree (AST) analysis to measure structural similarity. Building on these measures, we propose \underline{R}esidual \underline{R}isk \underline{S}coring (RRS), a unified framework that integrates embedding-based semantic similarity, localized AST-based structural similarity, and cross-model agreement to estimate residual risk in code. Our analysis shows that benign functions often remain highly similar to their vulnerable counterparts both semantically and structurally, indicating potential persistence of residual risk. We further find that approximately \textbf{61\%} of high-RRS C/C++ code pairs exhibit \textbf{13} distinct categories of residual issues (e.g., null pointer dereferences, unsafe memory allocation), validated using state-of-the-art static analysis tools including Cppcheck, Clang-Tidy, and Facebook-Infer. These results demonstrate that code-level similarity provides a practical signal for prioritizing post-patch inspection and residual risk assessment. By providing a quantitative measure of residual risk, RRS supports software security assurance through risk-informed post-patch prioritization and patch validation.

\end{abstract}


\vspace{-0.3cm}
\section{Introduction}
\vspace{-0.2cm}
Open-source software systems (e.g., Linux, Xserver, OpenSSL, Chrome) form the backbone of modern computing, where continuous development and widespread reuse demand high levels of robustness and reliability. To maintain this robustness, software is frequently updated through defect correction, feature enhancement, and, critically, vulnerability patching often on a daily basis in large-scale projects. In security-critical contexts, these patches are designed to eliminate reported flaws while preserving functional correctness and minimizing unintended side effects. However, even after patching, code often retains key semantic and structural traits of the original vulnerable implementation, allowing residual risks to persist. This uncertainty regarding patch effectiveness complicates software security assurance by limiting confidence in the risk-reduction achieved by a patch.\\
\indent In practice, developers tend to focus on quick, targeted fixes, prioritizing functionality and release constraints while avoiding extensive refactoring unless strictly required. Consequently, while such patches may be correct with respect to the identified vulnerability, they do not necessarily induce a substantial transformation of the underlying semantics or structure of the affected function. The rapid expansion of open-source software ecosystems further amplifies these challenges by increasing both code reuse and exposure to hidden residual vulnerabilities \cite{ZIO2016137}. Public repositories host millions of patches that are integrated and reused at scale, often without thoroughly checking whether the fix fully resolves all underlying logical and structural issues. Although vulnerability patches are released frequently, their effectiveness in fully mitigating underlying weaknesses remains uncertain \cite{Dijkstra1970NotesOS}. This uncertainty has motivated research into patch analysis, vulnerability repair automation, and residual risk estimation. For instance, multi-level code representations enable detection of subtle residual risks within patched code \cite{tang2023}. Similarly, hybrid approaches combining transformer-based representations with static analysis techniques have shown promise in identifying sensitive data exposures (CWE-200) that persist post-remediation \cite{katz2025}. Advances in large language models (LLMs) for automated patch generation highlight the potential of self-repairing software \cite{ye2025well}, yet these systems still struggle to reliably evaluate the semantic completeness of the fixes they propose.\\
\indent Despite significant progress in detection and repair, evaluating the efficacy of patches remains an open challenge \cite{wang2019different,vulrepair}. Traditional Static Application Security Testing (SAST) and learning-based vulnerability prediction frameworks offer partial insights into post-patch behavior, but they seldom quantify how function semantics change through patching \cite{croft2021empirical}. This gap limits our ability to reason about the degeneracy between vulnerable and benign representations an effect wherein benign code continues to resemble the vulnerable version closely, semantically or structurally. Function-level assessment offers a natural lens for analyzing these effects, facilitating fine-grained vulnerability characterization in languages such as C/C++ \cite{nguyen2024automated}. Understanding the semantic and structural deviations introduced by patches is essential for reliable vulnerability triage, automated repair validation, and model-based reasoning about software integrity \cite{xie2024unveiling}.\\
\indent From a broader reliability perspective, these issues echo the notion of \underline{\textit{residual}} \newline 
\underline{\textit{risk}}, the uncertainty that remains after exhaustive testing or patch deployment \cite{2018STAD}. Even when a testing campaign reveals no defects, the probability that future inputs might still uncover latent vulnerabilities remains strictly positive \cite{Dijkstra1970NotesOS}. If unaddressed, this residual risk can eventually manifest as zero-day vulnerabilities, which refer to previously unknown security flaws that are exploited before developers have the opportunity to identify, patch, or disclose them \cite{bilge2012before}. \\
\indent Recent work on dependency-aware risk estimation refines classical models by accounting for conditional dependencies among execution paths, resulting in more accurate residual risk bounds than traditional estimators such as Good–Turing \cite{lee2026dependency}. Together, these efforts raise the question of whether patched (benign) code undergoes sufficient semantic and structural transformation to meaningfully reduce security risk. Importantly, \textit{residual risk} does not imply patch incorrectness or immediate exploitability. Rather, it reflects the practical nature of software maintenance, where patches are often minimal and localized, addressing specific fault conditions while leaving broader logic unchanged. As a result, benign code may remain semantically and structurally close to its vulnerable counterpart, potentially retaining risky implementations patterns (e.g., null dereference, buffer overflow) and preserving related control flows, data dependencies that warrant further inspection. At the same time, minimal changes are often sufficient and desirable, as they prevent vulnerabilities while preserving system stability and avoiding unintended side effects. However, the relationship between code similarity and residual risk may vary across patch types, particularly those combining vulnerability fixes with feature additions.\\
\indent Parallel advances in semantic code analysis highlight the importance of measuring code similarity for tasks such as clone detection, code retrieval, and plagiarism detection \cite{song2024revisiting}. Embedding-based and structure-aware models further enable capturing underlying semantics across programming languages \cite{ebrahim2023}. However, these techniques are rarely applied to security patches, leaving unanswered the question of \textit{'whether semantic and structural similarity between vulnerable and benign code can serve as an indicator of residual risk'}. We evaluate the proposed framework on vulnerable--benign C/C++ function pairs from PrimeVul, a benchmarking dataset using multiple code language models and localized AST analysis. Our results show that benign functions frequently remain highly similar to their vulnerable counterparts, while static-analysis validation reveals that 61\% of high-RRS function pairs exhibit 13 categories of security-relevant residual issues. These findings suggest that representation-level similarity provides a practical signal for post-patch prioritization and software security assurance by helping identify patches that warrant further security inspection.

\noindent \textbf{Contributions.} This paper makes the following contributions:
\begin{enumerate}
\vspace{-0.2cm}
\item We propose Residual Risk Scoring (RRS), a framework that quantifies residual risk using semantic similarity, localized AST similarity, and cross-model agreement. To our knowledge, this is the first work to formalize residual risk through representation-level similarity between vulnerable and benign code.
\item We show that localized AST analysis more accurately captures patch-induced changes than global tree-based metrics, which can obscure fine-grained, security-relevant modifications.
\item We validate the practical utility of RRS through empirical analysis of high-risk benign functions, leveraging state-of-the-art C/C++ static analysis tools (Cppcheck, Clang-Tidy, Facebook Infer) and manual inspection to identify residual security issues.
\end{enumerate}

\vspace{-0.5cm}
\section{Motivation: An Illustrative Scenario}
\label{sec:back}

\begin{figure}[htbp]
\centering

\begin{minipage}[t]{0.30\linewidth}
    \centering
    \includegraphics[width=\linewidth]{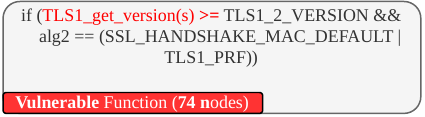}

    \vspace{0.3em} 

    \includegraphics[width=\linewidth]{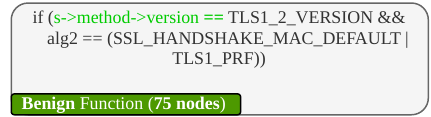}
\end{minipage}
\begin{minipage}{0.24\linewidth}
    \centering
    \includegraphics[width=\linewidth,height=3.2cm]{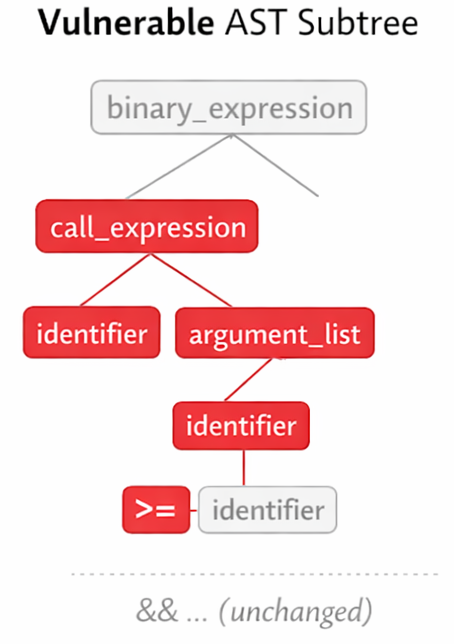}
\end{minipage}
\hspace{-0.3cm}
\begin{minipage}{0.24\linewidth}
    \centering
    \includegraphics[width=\linewidth,height=3.2cm]{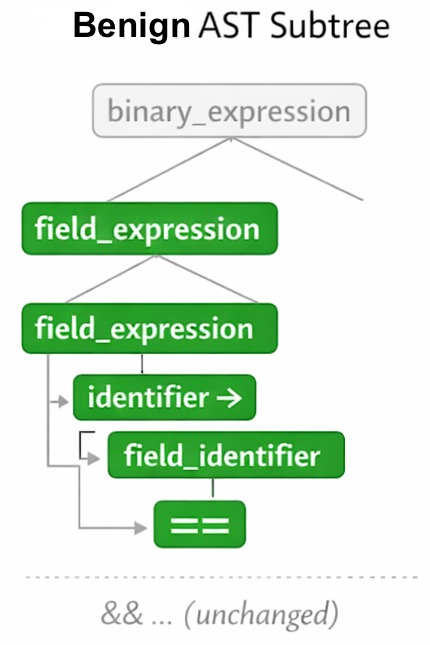}
\end{minipage}

\vspace{-0.4em}

\caption{Vulnerable and benign code (left) with corresponding AST subtrees (right).}
\label{fig:four_in_row}
\vspace{-0.6cm}
\end{figure}

Figure \ref{fig:four_in_row} shows a minimal syntactic change between a vulnerable and benign function pair drawn from a real-world vulnerability patch from the PrimeVul \cite{primeVul} and analyzed in its original form, where a version check is refactored from a function call–based comparison ($\geq$) to a direct field comparison ($==$). Although this appears as a minor token-level modification, the corresponding ASTs differ only in a small subtree, replacing a \texttt{call\_expression} with a \texttt{field\_expression} while preserving the overall structure. The functions contain 74 and 75 AST nodes, respectively, indicating that most of the structure remains unchanged.

\vspace{-0.5cm}

\noindent Despite this localized change, global tree edit distance (TED) which measures the minimum number of node insertions, deletions, and substitutions required to transform one AST into another reports a much larger edit cost. While the actual modification involves only $\sim$10 node-level edits (\underline{5 deletions} in red and \underline{5 insertions} in green), yielding a localized TED similarity of 0.91, the global TED reports 60 operations (normalized similarity 0.20). This inflation occurs because global TED treats modified subtrees as bulk replacements, propagating edits across the hierarchy. Semantically, the function remains largely unchanged (embedding similarity 0.96), highlighting a mismatch between syntactic magnitude and semantic impact. Thus, global TED may overestimate structural change, while embeddings alone may overlook residual risk. Localized TED better captures patch-relevant changes and aligns more closely with semantic similarity.

From a structural perspective, AST-based methods such as TED provide richer insight than simpler measures (e.g., Jaccard similarity) by capturing hierarchical transformations. However, structural and semantic signals are complementary and must be considered jointly. Importantly, existing vulnerability workflows often assume a binary transition from vulnerable to benign after patching \cite{Grape,GraphSPD}. This overlooks the representational continuity between versions. High similarity does not imply developer error, but rather reflects narrowly scoped fixes. From a security perspective, such proximity may warrant further inspection to ensure that remediation extends beyond the immediate condition. Recent advances in code language models (e.g., CodeBERT, CodeT5) enable semantic comparison via embeddings, which effectively capture functional behavior across tasks \cite{ebrahim2023,katz2025}. However, existing datasets and evaluation frameworks rarely combine semantic and structural signals to assess post-patch risk. High embedding similarity and minimal structural change together suggest the possibility of residual risk, where benign code remains closely aligned with its vulnerable predecessor and may require additional scrutiny.

\vspace{-0.4cm}
\section{Residual Risk Assessment Approach}
\begin{figure*}[htbp]
    \centering
    \includegraphics[width=0.8\linewidth]{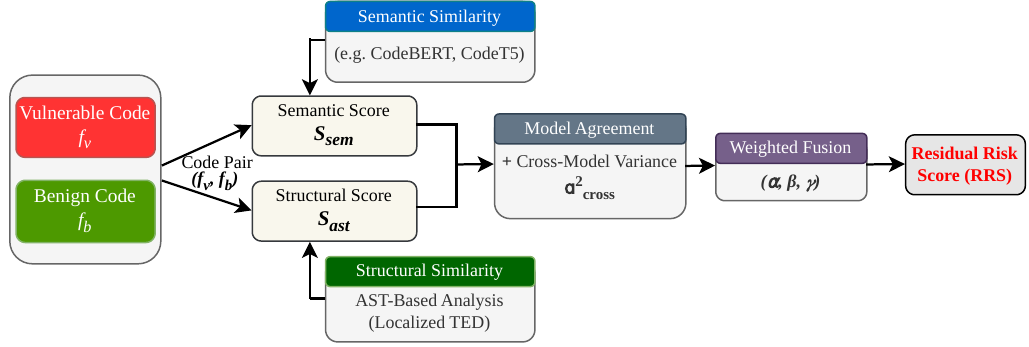}
    \caption{Residual risk assessment pipeline combining semantic similarity ($S_{\text{sem}}$), structural similarity ($S_{\text{ast}}$), and cross-model ($\sigma^2_{\text{cross}}$) agreement to compute RRS.}

    \label{fig:rrs}
\end{figure*}

\vspace{-0.2cm}
\label{sec:approach}
\noindent Our approach is motivated by the observation that vulnerability patches often introduce localized fixes while preserving large portions of the original code structure and behavior. Given a vulnerable–benign function pair $(f_v, f_b)$, we encode both functions using multiple pretrained code language models to capture their semantic representations across diverse model architectures. To capture structural changes, we compute AST-based similarity focusing on \textit{localized} structural differences that reflect minimal modified regions introduced by the patch. Additionally, we quantify cross-model variance to assess the consistency of semantic similarity across different models, providing an additional signal of representational uncertainty. As illustrated in Figure \ref{fig:rrs}, these components are integrated to derive a Residual Risk Score (RRS), combining semantic similarity, localized structural similarity, and cross-model agreement to estimate the residual risk associated with benign (patched) code.

\vspace{-0.4cm}
\noindent Table \ref{tab:code_lms} summarizes the code language models (Code LMs) utilized in this study, covering a diverse set of architectures and model scales. Specifically, we include encoder-based models such as CodeBERT, UniXCoder, and GraphCodeBERT, each with approximately 125M parameters, where GraphCodeBERT further incorporates data-flow graph (DFG) information to capture richer structural dependencies in code. In addition, we consider encoder–decoder architectures, namely CodeT5-base (220M) and CodeT5-770M (770M), which enable more expressive sequence-to-sequence representations. All models are employed in a fine-tuning setting to adapt their pretrained representations to our analysis task. This selection allows us to systematically evaluate the impact of architectural design (encoder vs. encoder–decoder), structural awareness (e.g., DFG), and model scale on capturing semantic relationships between vulnerable and benign code.
Each model encodes vulnerable and benign functions into dense embeddings. For each vulnerable–benign function pair ($f_v$, $f_b$), we compute:
\[
X_m = E_m(f_v), \quad Y_m = E_m(f_b)
\]
where $E_m(.)$ denotes the embedding function of model $m$.

\vspace{-0.8cm}
\begin{table}[htbp]
\centering
\scriptsize
\caption{The code language models (Code LMs) analyzed in this study.}
\label{tab:code_lms}
\begin{tabular}{lccc}
\hline
\textbf{Model} & \textbf{Parameters} & \textbf{Type} & \textbf{Usage} \\
\hline

CodeBERT \cite{codebert}   & 125M   & Encoder          & Fine-tune \\
UniXCoder \cite{unixcoder} & 125M   & Encoder          & Fine-tune \\
GraphCodeBERT \cite{graphcodebert} & 125M   & Encoder+DFG          & Fine-tune \\
CodeT5-base \cite{codet5}       & 220M    & Enc--Dec & Fine-tune \\
CodeT5-770m \cite{codet5plus}& 770M    & Enc--Dec & Fine-tune \\

\hline
\end{tabular}
\end{table}
\vspace{-1cm}
\subsection{Embeddings Representation Similarity}
\noindent To quantify semantic preservation after patching, we compute cosine similarity $S_{m}$ between $f_v$ and $f_b$:

\begin{equation}
S^{\mathrm{cos}}_{m}(f_v, f_b) 
= 
\frac{X_m \cdot Y_m}{\|X_m\| \|Y_m\|}
\end{equation}

\noindent Here, $X_m$ and $Y_m$ denote the vector embeddings of the vulnerable function $f_v$ and the benign function $f_b$, respectively, obtained using model $m$. $\|\cdot\|$ denotes the Euclidean norm. This metric reflects how closely the benign function remains to the vulnerable version in the learned representation space. High similarity indicates that the functional behavior encoded by the model has not significantly changed, which may imply incomplete mitigation.

To justify the choice of cosine similarity, Table \ref{tab:embedding_metrics} compares multiple distance and similarity metrics computed over vulnerable--benign function pairs from primevul \cite{primeVul} dataset using model $m =$ CodeBERT as a representative example. We observe that cosine similarity consistently produces highly stable and normalized scores (mean: 0.9988, standard deviation: 0.0036), where values closer to 1 indicate greater similarity, effectively capturing semantic proximity while remaining invariant to embedding magnitude. In contrast, metrics such as dot product are sensitive to vector scale, while distance-based measures (e.g., Euclidean, Manhattan, Chebyshev) have a lower bound of 0, with larger values indicating greater dissimilarity, but lack direct interpretability in terms of semantic equivalence.

\vspace{-0.7cm}

\begin{table}[htbp]
\centering
\scriptsize
\caption{Distance metrics and average similarity across vulnerable--benign function pairs from PrimeVul (Model $m$ = CodeBERT).}
\label{tab:embedding_metrics}
\begin{tabular}{l| c| c| c| c}
\hline
\textbf{Metric} & \textbf{Formula} & \textbf{Mean} & \textbf{Median} & \textbf{Std Dev}\\
\hline
Cosine \textbf{(Baseline)} &
$\displaystyle \frac{\mathbf{X_m} \cdot \mathbf{Y_m}}{\|\mathbf{X_m}\|_2 \|\mathbf{Y_m}\|_2}$ &
\cellcolor{gray!30} 0.9988 & \cellcolor{gray!30} 0.9998 & \cellcolor{gray!30} 0.0036 \\
\hline
Dot &
$\displaystyle \mathbf{X_m} \cdot \mathbf{Y_m}$ &  338.75 & 337.39 & 13.33\\
\hline
Manhattan (L1) &
$\displaystyle \|\mathbf{X_m} - \mathbf{Y_m}\|_1$ & 9.96 & 6.87 & 12.26\\
\hline
Euclidean (L2) &
$\displaystyle \|\mathbf{X_m} - \mathbf{Y_m}\|_2$ & 0.56 & 0.34 & 0.75\\
\hline
Chebyshev (L$\infty$) &
$\displaystyle \max_i |X_{m}^i - Y_{m}^i|$ & 0.24 & 0.08 & 0.42 \\
\hline
\end{tabular}
\end{table}

\noindent Based on these observations, we adopt \textit{baseline cosine similarity} as our primary semantic similarity measure due to its robustness, bounded range, and consistency across samples. Prior studies have also employed cosine similarity over learned code embeddings as an effective measure of semantic relatedness in tasks such as plagiarism detection and vulnerability analysis \cite{ebrahim2023,nikiema2025small}. We note that similar trends are observed across other code language models considered in this study, indicating that cosine similarity provides a reliable and model-agnostic measure for comparing vulnerable and benign code representations.

\subsection{Localized Abstract Syntax Tree Analysis}
\noindent Semantic similarity alone is insufficient to capture fine-grained structural changes introduced by patches. To address this, we analyze structural similarity using Tree-sitter–derived ASTs. Let $T_v$ and $T_b$ denote the ASTs of vulnerable and benign functions. Let $lTED_{local}(T_v,T_b)$ denote the localized tree edit distance computed over the modified AST region. We compute structural similarity as:


\begin{equation}
S_{\mathrm{AST}}(f_v, f_b)
=
1 - \frac{TED_{local}(T_v, T_b)}{\max(|T_v|, |T_b|)}
\end{equation}

\noindent
where $|T_v|$ and $|T_b|$ denote the number of nodes in the localized AST regions of the vulnerable and benign functions, respectively. Higher values indicate smaller structural changes. We interpret this as \textit{structural residuality}, defined as: $R_{\mathrm{struct}} = S_{\mathrm{AST}}$.

\noindent \textbf{Limitations of Global Structural Metrics.} A key design choice in our approach is the use of \textit{localized} AST similarity rather than global tree-based metrics. Vulnerability patches are typically minimal and localized, affecting only small regions while leaving most of the structure unchanged. As a result, global metrics such as NTED dilute these changes by averaging over large unchanged regions, often masking subtle but security-critical modifications. Prior work on AST differencing similarly shows that global comparisons capture overall syntactic variation but fail to reflect localized behavioral changes \cite{song2024revisiting,falleri2014fine,cheng2022path}. In our analysis, we retain cases where global TED yields very low or degenerate similarity scores, not as noise, but as evidence of this limitation.

To address this, we focus on modified subtrees corresponding to patches and compute Localized TED Similarity (LTS). Table \ref{tab:ast-similarity} compares alternative structural measures. NTED produces substantially lower similarity scores (mean 0.27, median 0.34) due to its sensitivity to full-tree variation, whereas Jaccard Similarity (mean 0.94, median 0.97) ignores hierarchy and ordering. Alignment Similarity captures node correspondence across aligned subtrees, while LTS provides a balanced view of structural preservation (mean 0.82, median 0.93) by focusing on patch-relevant changes. We therefore adopt LTS as the structural component of RRS.

\vspace{-0.5cm}
\begin{table}[htbp]
\centering
\scriptsize
\caption{Several metrics of AST structural similarity for vulnerable--benign function pairs.}
\label{tab:ast-similarity}
\begin{tabular}{
>{\raggedright\arraybackslash}p{2.5cm}| >{\centering\arraybackslash}p{7cm}| >{\centering\arraybackslash}p{1cm}|c}
\hline
\textbf{Metrics} & \textbf{Formula} & \textbf{Mean} & \textbf{Median} \\
\hline

LTS (\textbf{Utilized}) &
$\displaystyle lTED_{local}(T_1,T_2) = 1- \frac{TED_{local}(T_1,T_2)}{\max(|T_1|, |T_2|)}$ &
\cellcolor{gray!30} 0.82 & \cellcolor{gray!30} 0.93 \\
\hline

NTED Similarity (\textbf{Baseline})  &
$\displaystyle TED_{norm}(T_1,T_2) = 1- \frac{TED(T_1,T_2)}{\max(|T_1|, |T_2|)}$ &
\underline{0.27} & \underline{0.34} \\

\hline
JS &
$S_{\mathrm{AST}}(f_v, f_b) = \frac{|N(f_v) \cap N(f_b)|}{|N(f_v) \cup N(f_b)|}$ &
0.94 & 0.97 \\
\hline
AS (\textbf{Custom}) &
$\displaystyle AlignSim(f_v, f_b) = \frac{\text{matches}(f_v, f_b)}{|N(f_v) \cup N(f_b)|_{\text{alignment}}}$ &
0.75 & 0.81 \\
\hline
\end{tabular}

\vspace{0.3em}
\scriptsize{
\textbf{NTED} = Normalized Tree Edit Distance (Global); 
\textbf{LTS} = Localized TED Similarity; 
\textbf{JS} = Jaccard Similarity; 
\textbf{AS} = Alignment Similarity. 
}

\end{table}


\subsection{Cross-Model Semantic Agreement}

\noindent
While individual models may produce noisy or biased similarity estimates, consistent agreement across multiple models provides stronger evidence of semantic preservation. To capture this, we compute the mean semantic similarity:

\begin{equation}
\bar{S}_{\mathrm{sem}}(f_v, f_b)
=
\frac{1}{M}
\sum_{m=1}^{M}
S^{\mathrm{cos}}_{m}(f_v, f_b)
\end{equation}

\noindent We further quantify agreement using cross-model variance:

\begin{equation}
\sigma_{\mathrm{cross}}^{2}
=
\frac{1}{M}
\sum_{m=1}^{M}
\left(
S^{\mathrm{cos}}_{m}(f_v, f_b)
-
\bar{S}_{\mathrm{sem}}(f_v, f_b)
\right)^{2}
\end{equation}

\noindent
Low variance indicates strong agreement among models. We normalize this into an agreement score, $C_{\mathrm{agree}}$:

\begin{equation}
C_{\mathrm{agree}} = 1 - \frac{\sigma_{\mathrm{cross}}}{\sigma_{\max}}
\end{equation}

\noindent
where $\sigma_{\max}$ is the maximum observed variance across all pairs. This formulation bounds $C_{\mathrm{agree}} \in [0,1]$, where values close to 1 indicate strong agreement (low variance) across models, and values close to 0 indicate high disagreement.

\subsection{Residual Risk Scoring (RRS)}

\noindent We define residual risk as a composite signal that emerges when (1). semantic similarity remains high, (2). structural changes are limited, and (3). multiple models consistently agree. These three conditions jointly indicate that the benign code remains close to its vulnerable counterpart in both representation and structure. The final Residual Risk Score (RRS) is computed as:
\begin{equation}
\mathrm{\textbf{RRS}}(f_v, f_b) 
= \alpha \cdot \bar{S}_{\text{sem}} 
+ \beta \cdot R_{\text{struct}} 
+ \gamma \cdot C_{\text{agree}}
\end{equation}

\noindent
where $\alpha, \beta, \gamma \in [0,1]$ and $\alpha + \beta + \gamma = 1$. As an example, we set: $\alpha = 0.5, \beta = 0.3, \gamma = 0.2$. Importantly, semantic similarity ($\bar{S}_{\text{sem}}$) captures functional equivalence between vulnerable and benign code, while structural similarity ($R_{\text{struct}}$) reflects localized code transformations introduced by patches. Cross-model agreement ($C_{\text{agree}}$) provides complementary evidence by measuring the consistency of similarity scores across different code language models, thereby capturing representational robustness. To integrate these signals, we adopt a weighted formulation where $\alpha$ and $\beta$ control the contribution of semantic and structural similarity, respectively, while $\gamma$ captures cross-model agreement. The weighting coefficients $(\alpha, \beta, \gamma)$ balance the contribution of these heterogeneous signals within a unified framework, rather than serving as a normalization mechanism, as all signals are normalized to a common range prior to aggregation.

To ensure that our analysis is not dependent on specific weight assignments, we conduct a sensitivity analysis by varying $\alpha$ and $\beta$ over $[0,1]$ under the constraint $\alpha + \beta + \gamma = 1$ (Section \ref{sec:results-rq2}). We fix $\gamma = 0.2$ as a moderate auxiliary weight, since cross-model agreement serves as a consistency signal rather than primary evidence of residual risk. Lower values underutilize agreement, while higher values risk overshadowing the semantic and structural signals. This design enables us to analyze their trade-off without allowing agreement to dominate the overall score.

\vspace{-0.5cm}
\section{Experimental Design}
\label{sec:ex-design}

\vspace{-0.3cm}
\subsection{Research Questions}

\begin{tcolorbox}[
    colframe=black,
    colback=white,
    boxsep=2pt,     
    left=4pt,
    right=4pt,
    top=2pt,
    bottom=2pt,
    boxrule=1pt,
]
\textbf{RQ1:} How closely do benign functions preserve the semantic and structural properties of their vulnerable predecessors?
\end{tcolorbox}

\noindent This research question examines how vulnerability patches affect code representations. We measure embedding-based semantic similarity and localized AST-based structural similarity between vulnerable and benign functions to quantify residual similarity after patching. Results are presented in Section~\ref{sec:results-rq1}.

\begin{tcolorbox}[
    colframe=black,
    colback=white,
    boxsep=2pt,     
    left=4pt,
    right=4pt,
    top=2pt,
    bottom=2pt,
    boxrule=1pt,
]
\textbf{RQ2:} How effectively does joint embedding and AST-based similarity estimate residual risk in benign code?
\end{tcolorbox}

\noindent We evaluate whether combining semantic similarity, structural similarity, and cross-model agreement through RRS provides a robust characterization of residual risk. We further assess whether this multi-signal approach better characterizes residual risk than any single metric. Results in Section~\ref{sec:results-rq2} assess the effectiveness and stability of the proposed RRS.

\begin{tcolorbox}[
    colframe=black,
    colback=white,
    boxsep=2pt,     
    left=4pt,
    right=4pt,
    top=2pt,
    bottom=2pt,
    boxrule=1pt,
]
\textbf{RQ3:} To what extent do high-RRS benign functions exhibit potential security-relevant issues identified through state-of-the-art static analysis tools?
\end{tcolorbox}
\noindent Building on RQ1 and RQ2, this research question evaluates whether high residual risk scores correspond to observable security-relevant issues in practice. Specifically, we assess whether benign functions identified as high-risk by RRS exhibit potential security-relevant issues under state-of-the-art static analysis tools. The results are presented in Section \ref{sec:results-rq3}.

\vspace{-0.3cm}
\subsection{Dataset \& Implementation Setup}

We use PrimeVul \cite{primeVul}, a widely adopted benchmark dataset of real-world C/C++ software vulnerabilities that provides vulnerable functions together with their corresponding benign versions and vulnerability annotations. While other vulnerability datasets are available, many are primarily designed for vulnerability detection and classification (e.g., Devign \cite{Devign}, ReVeal \cite{chakraborty2020deep}). PrimeVul was selected because it provides explicitly paired vulnerable--benign functions, which are essential for measuring semantic and structural similarity before and after patching. For our study, we use all 3,789 vulnerable–benign function pairs available in the publicly released paired subset of PrimeVul, resulting in a total of 7,578 functions. The dataset covers diverse vulnerability types, patch sizes, and code contexts drawn from real-world open-source projects. Each pair is treated as a unit of analysis for computing embedding-based similarity and structural changes. This dataset forms the foundation of our RRS framework, enabling quantification of post-patch similarity and identification of functions that may require further inspection using static analysis tools.\\
\indent For representation, we use pretrained code language models listed in Table \ref{tab:code_lms}, obtained through the Hugging Face repository \cite{huggingface}, to generate function embeddings. For each function, source code is tokenized using the corresponding model tokenizer and passed through the model; mean pooling over the final hidden states is then applied to obtain a fixed-length embedding. Structural analysis is performed using ASTs constructed via the Tree-sitter framework \cite{tree_sitter}, which provides accurate, language-aware parsing for C/C++ code. Since PrimeVul provides raw C/C++ source code rather than pre-generated ASTs, we independently generate ASTs using the official Tree-sitter C and C++ grammars.

\vspace{-0.4cm}
\section{Results}
\label{sec:results}
\subsection{RQ1: Semantic \& Structural Properties Preservation}
\label{sec:results-rq1}

\noindent \textbf{Embedding-Based Semantic Similarity.} Table \ref{tab:embedding-stats} shows consistently high similarity scores across all code language models (typically 0.97--0.99), indicating strong preservation of semantic representations after patching. CodeBERT and GraphCodeBERT exhibit near-perfect similarity (means of \underline{0.9988} and \underline{0.9970}, medians $\approx$ 1.0), with very low variance. CodeT5 variants show similarly high values, while UniXCoder exhibits slightly higher variability but maintains a high median (0.9929), suggesting that most function pairs remain semantically close.

\vspace{-0.7cm}
\begin{table}[htbp]
\centering
\scriptsize
\caption{Embedding cosine similarity statistics for different code LM models.}
\label{tab:embedding-stats}
\begin{tabular}{l|c|c|c}
\hline
\textbf{Model} 
& \multicolumn{3}{c}{\textbf{Embedding}} \\
\cline{2-4}
& \textbf{Mean} & \textbf{Median} & \textbf{Std Dev} \\
\hline
CodeBERT        & \cellcolor{gray!30} 0.9988 & \cellcolor{gray!30} 0.9998 & \cellcolor{gray!30} 0.0036 \\
\cline{2-4}
UniXCoder       & 0.9767 & 0.9929 & 0.0442 \\
\cline{2-4}
GraphCodeBERT   & \cellcolor{gray!30} 0.9970 & \cellcolor{gray!30} 0.9995 & \cellcolor{gray!30} 0.0087 \\
\cline{2-4}
CodeT5-base     & 0.9834 & 0.9978 & 0.0396 \\
\cline{2-4}
CodeT5-770m     & 0.9867 & 0.9979 & 0.0228 \\
\cline{2-4}
\hline
\end{tabular}
\end{table}

\begin{figure}[htbp]
\centering

\begin{subfigure}{0.48\linewidth}
    \centering
    \includegraphics[width=\linewidth]{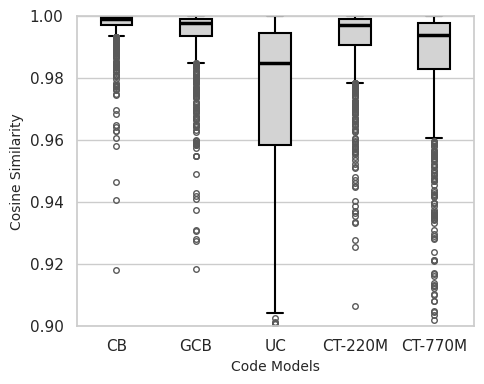}
    \caption{Embedding consistency across models.}
    \label{fig:code-lm cosine}
\end{subfigure}
\hfill
\begin{subfigure}{0.48\linewidth}
    \centering
    \includegraphics[width=\linewidth]{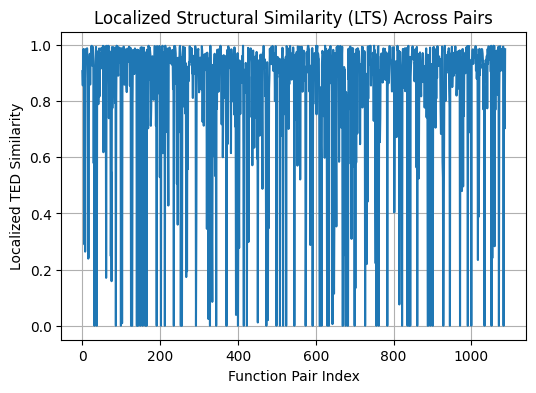}
    \caption{Distribution of Localized TED Similarity (LTS).}
    \label{fig:lts_plot}
\end{subfigure}

\caption{Semantic and structural similarity analysis across vulnerable--benign function pairs.}
\label{fig:combined_analysis}

\end{figure}

\noindent \textbf{Implications of High Semantic Similarity.} These results reflect real-world patch behavior: fixes are typically minimal and localized (e.g., boundary checks), preserving overall functionality. Consequently, embedding-based models, capturing global semantics, produce highly similar representations for vulnerable and benign (Fig. \ref{fig:code-lm cosine}). Importantly, this does not indicate model weakness or poor patch quality. Instead, it highlights a fundamental challenge: \textit{semantic similarity alone cannot capture subtle, security-critical changes}, motivating the need for complementary structural analysis.

\noindent \textbf{Localized Structural Similarity (LTS) Analysis.} Fig. \ref{fig:lts_plot} shows that most function pairs cluster in the high-similarity range [0.8–1.0], indicating that patches typically introduce limited structural modifications while preserving overall structure.
However, occasional drops in similarity reveal cases with substantial structural changes (e.g., control flow). The distribution is therefore skewed: a high median (0.93) reflects dominant localized edits, while a lower mean (0.82) is influenced by outliers. Note that, due to computational constraints, we limit RRS computation to function pairs with AST size $\leq$ 350 nodes ($\approx$1088 pairs).

\vspace{-0.5cm}



\vspace{-0.7cm}
\begin{tcolorbox}[
    colframe=black,
    colback=gray!30,    
    boxsep=2pt,     
    left=4pt,
    right=4pt,
    top=2pt,
    bottom=2pt,
    boxrule=0.5pt,
]
\textbf{Answer to RQ1:} Benign functions remain highly similar to their vulnerable counterparts across multiple code language models, with near-unity cosine similarity and low variance, indicating that vulnerability patches often preserve substantial semantic and structural properties of the original code.
\end{tcolorbox}

\vspace{-0.3cm}
\subsection{RQ2: Residual Risk Estimation}
\label{sec:results-rq2}
\noindent \textbf{Quadrant-Based Risk Characterization.} To evaluate residual risk, we partition function pairs into four quadrants based on embedding ($E$) and structural ($A$) similarity (Table \ref{tab:quadrant-count}), using median-based thresholds ($E=0.994$, $A=0.927$).
where \textit{Quadrant I (29.5\%)}: High $E$, High $A$ → minimal change, strong candidates for residual risk., \textit{Quadrant II (20.5\%)}: High $E$, Low $A$ → structural change missed by embeddings, \textit{Quadrant III (20.6\%)}: Low $E$, High $A$ → cosmetic/lexical changes and \textit{Quadrant IV (29.4\%)}: Low $E$, Low $A$ → substantial transformation, likely mitigated.\\
\noindent \textbf{Limitations of Single-Signal Analysis.} This distribution highlights that relying on a single signal is insufficient: (a). Embeddings may \textit{overestimate equivalence} (Quadrant II) (b). Structural metrics may \textit{miss semantic preservation} (Quadrant III). These mismatches reveal blind spots in isolated representations.\newline
\textbf{Multi-Signal Advantage (RRS).} By jointly combining signals, our approach provides a more reliable estimate of residual risk. Quadrant I emerges as a critical region where minimal observable change may mask persistent vulnerabilities.\newline
\textbf{Robustness to Weight Sensitivity.} To validate robustness, we analyze RRS under varying semantic ($\alpha$) and structural ($\beta$) weight configurations in Fig. \ref{fig:sensitivity} (additional configurations are provided in Appendix A, Fig. \ref{fig:d-e}). Across all settings, the distribution and relative positioning of function pairs remain stable. High-risk clusters (Quadrant I) consistently persist, indicating that risk identification is not sensitive to parameter selection. Although RRS values vary slightly with weighting, the relative ranking of function pairs remains largely unchanged, confirming that high-risk candidates reflect inherent transformation properties rather than parameter artifacts.\newline
\noindent \textbf{Cross-Signal Interaction Analysis.} Fig. \ref{fig:multisignal-bars} in Appendix A, further illustrates the interaction between signals. Embedding similarity and cross-model agreement remain consistently high, reflecting stable global semantics, while localized AST similarity varies more, capturing fine-grained structural changes. Notably, several cases exhibit high semantic similarity but lower structural similarity, reinforcing that embedding-based representations alone may overlook meaningful structural modifications. These findings confirm that combining semantic, structural, and cross-model signals yields more accurate and robust residual risk estimation.\\

\vspace{-1.4cm}
\begin{table}[htbp]
\centering
\scriptsize
\caption{Distribution and impact of code pair types based on semantic and structural representation changes ($\approx$1088 function pairs).}
\label{tab:quadrant-count}
\begin{tabular}{>{\centering\arraybackslash}p{1.8cm}|
                >{\centering\arraybackslash}p{3cm}|
                >{\centering\arraybackslash}p{0.8cm}|
                >{\centering\arraybackslash}p{0.8cm}|
                >{\centering\arraybackslash}p{5.2cm}}
\hline
\textbf{Quadrant} & \textbf{RRS} & \textbf{Pairs} & \textbf{(\%)} & \textbf{Behavioral Interpretation} \\
\hline

I ($E_H$, $A_H$) 
& \cellcolor{gray!30} $\geq$ 97\% 
& \cellcolor{gray!30} 321 
& \cellcolor{gray!30} 29.5\%  
& \cellcolor{gray!30} Minor change, likely \textit{residual risk} \\
 \hline
II ($E_H$, $A_L$)  
& $0.94\% \leq$ RRS $<$ 97\% 
& 223 
& 20.5\% 
& Structural change, embeddings blind spot \\

\hline
III ($E_L$, $A_H$) 
& $0.90\% \leq$ RRS $<$ 94\% 
& 224  
& 20.6\%  
& Cosmetic vs.\ lexical change \\

\hline
IV ($E_L$, $A_L$)  
& $<$ 90\% 
& 320 
& 29.4\% 
& Major change, likely mitigated \\

\hline
\end{tabular}

\vspace{2pt}
\scriptsize{
\textbf{$E_H$} = High Embedding Similarity; 
\textbf{$A_H$} = High Structural Similarity; 
\textbf{$E_L$} = Low Embedding Similarity; 
\textbf{$A_L$} = Low Structural Similarity.
}
\end{table}
\vspace{-0.8cm}
\begin{tcolorbox}[
    colframe=black,
    colback=gray!30,    
    boxsep=2pt,     
    left=4pt,
    right=4pt,
    top=2pt,
    bottom=2pt,
    boxrule=0.5pt,
]

\textbf{Answer to RQ2:} 
Joint semantic and structural similarity provides a stable estimate of residual risk by capturing complementary aspects of code transformation. High similarity across both signals identifies function pairs with minimal observable changes and higher \textit{residual risk} likelihood, while consistent results across weighting configurations demonstrate the robustness of RRS.
\end{tcolorbox}

\begin{figure*}[t]
\centering

\begin{subfigure}[t]{0.30\textwidth}
\centering
\includegraphics[width=\linewidth,height=3.5cm]{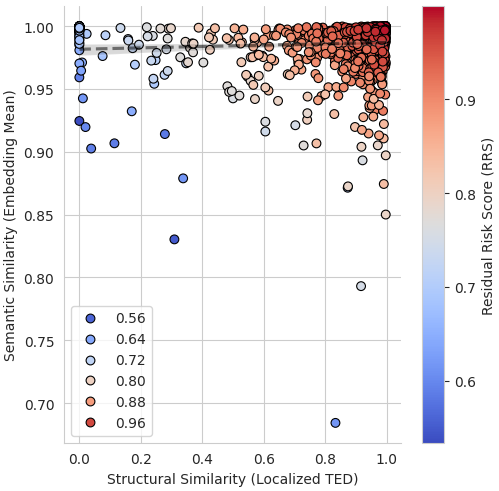}
\caption{$\alpha=0.5, \beta=0.3$}
\end{subfigure}
\hfill
\begin{subfigure}[t]{0.30\textwidth}
\centering
\includegraphics[width=\linewidth,height=3.5cm]{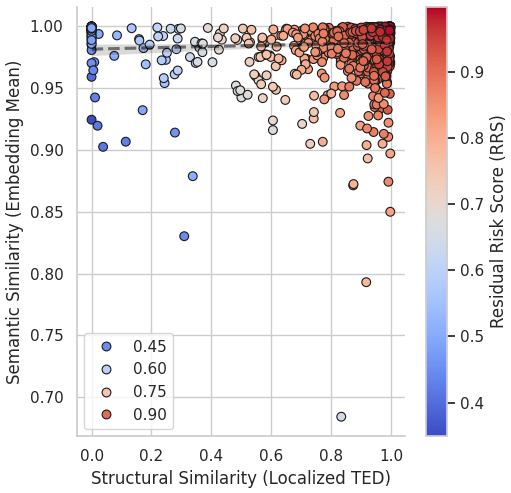}
\caption{$\alpha=0.3, \beta=0.5$}
\end{subfigure}
\hfill
\begin{subfigure}[t]{0.30\textwidth}
\centering
\includegraphics[width=\linewidth,height=3.5cm]{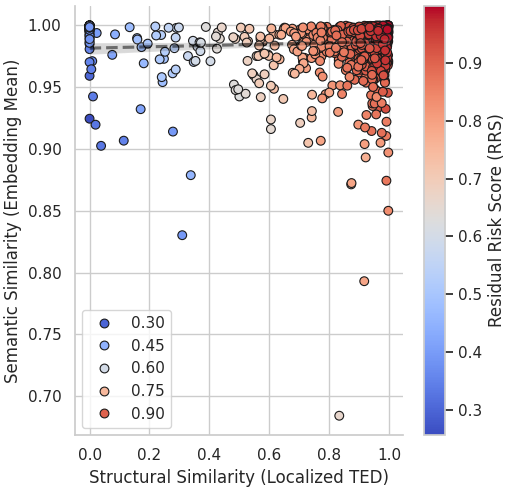}
\caption{$\alpha=0.4, \beta=0.4$}
\label{fig:sub5}
\end{subfigure}

\caption{Semantic vs. structural similarity for vulnerable--benign pairs under different $(\alpha,\beta)$ configurations ($\gamma=0.2$).}
\label{fig:sensitivity}

\end{figure*}

\vspace{-0.5cm}
\subsection{RQ3: State-of-the-art Static Analysis Tools}
\noindent To assess whether high residual risk corresponds to practical security concerns, we analyze high-RRS benign functions using state-of-the-art static analysis tools. We focus detailed validation on Quadrant I (29.5\% of all pairs), which represents cases with both high semantic and structural similarity and forms the primary candidate space for residual risk. While residual issues are evaluated across all quadrants, our goal is to determine whether RRS effectively prioritizes the most likely high-risk cases. We evaluate top-ranked high-RRS function pairs using \texttt{Cppcheck} \cite{cppcheck}, \texttt{Clang-Tidy} \cite{clang-tidy}, and \texttt{Facebook Infer} \cite{infer}. \textit{Cppcheck} provides lightweight detection of undefined behavior and common errors, \textit{Clang-Tidy} performs rule-based and AST-level checks, and \textit{Infer} developed by Meta \cite{meta} applies interprocedural, path-sensitive analysis to detect issues such as null dereferences and memory errors. Importantly, RQ3 focuses on identifying residual security-relevant issues in benign code rather than determining whether detected issues correspond to the original vulnerability. Our objective is to evaluate the persistence of residual risk following patching, irrespective of the specific vulnerability category.\\
\indent Our results show that approximately \textbf{61\%} of high-RRS functions are flagged by at least one tool for security-relevant issues, including memory safety violations, boundary errors, and improper error handling. Among these, \textbf{43\%} are flagged by two tools, and \textbf{27\%} by all three, indicating stronger agreement. Conversely, \textbf{39\%} produce no warnings, suggesting that not all high-similarity cases correspond to detectable issues. These findings indicate that RRS captures meaningful signals related to real security concerns rather than merely reflecting representational similarity. Partial disagreement across tools further highlights their complementary detection capabilities and the importance of multi-tool evaluation. Table \ref{tab:residual-issues-Q} in Appendix A, further shows that Quadrant I exhibits the highest proportion of residual issues identified by static analysis, substantially exceeding other quadrants providing empirical support for the relationship between high RRS and elevated residual risk.

\begin{table*}[t]
\centering
\scriptsize
\setlength{\tabcolsep}{3pt}
\caption{Representative validation outcomes from high-RRS function pairs.}
\label{tab:validation_results}

\begin{tabular}{l|>{\centering\arraybackslash}p{2cm}|>{\centering\arraybackslash}p{2cm}|>{\centering\arraybackslash}p{2cm}|c|c|c|c}
\hline
\textbf{Cases} & \textbf{Cppcheck} \cite{cppcheck} & \textbf{Clang-tidy} \cite{clang-tidy} & \textbf{Infer} \cite{infer} & \textbf{Agr.} & \textbf{R.I.} & \textbf{CVE} & \textbf{M.V.}\\
\hline

C1  & Uninit. var & --  & Uninit. var & $\LEFTcircle$ & $\checkmark$ & 2024-32878 & Y\\
C2  & -- & Dead store & Dead store & $\LEFTcircle$ & $\checkmark$ & 2023-53024 & Y\\
C3  & Unsafe realloc & Unsafe realloc & Dead store & $\LEFTcircle$ & $\checkmark$ & 2023-37573 & Y\\
C4  & Null ptr & -- & Null ptr & $\LEFTcircle$ & $\checkmark$ & 2025-29886 & Y\\
C5  & Invalid ptr & Invalid ptr & Invalid ptr & \ding{108} & $\checkmark$ & 2025-21919 & Y\\
C6  & Resource leak & Dead store & Resource leak & $\LEFTcircle$ & $\checkmark$ & 2024-31735 & Y\\
C7  & -- & Mem. leak & Mem. leak & $\LEFTcircle$ & $\checkmark$ & 2024-31735 & Y\\
C8  & Uninit. val & Uninit. val & Uninit. val & \ding{108} & $\checkmark$ & 2024-47540 & Y\\
C9  & -- & Return misuse & -- & $\LEFTcircle$ & $\checkmark$ & 2024-45337 & Y\\
C10 & Null ptr & Dead store & Null ptr & $\LEFTcircle$ & $\checkmark$ & 2025-38562 & Y\\
C11 & -- & Missing default & -- & $\LEFTcircle$ & $\checkmark$ & 2026-33064 & Y\\
C12 & -- & -- & Null ptr & $\LEFTcircle$ & $\checkmark$ & 2024-41130 & Y\\
C13 & -- & Int. truncation & -- & $\LEFTcircle$ & $\checkmark$ & 2022-34169 & Y\\
C14 & -- & Missing default & Uninit. val & $\LEFTcircle$ & $\checkmark$ & 2024-32878 & Y\\
C15 & Alloc overflow & Alloc overflow & Alloc overflow & \ding{108} & $\checkmark$ & 2025-13601 & Y\\

\hline
\end{tabular}

\vspace{0.2em}
\scriptsize{
\textbf{Agr.} = Agreement; \textbf{R.I.} = Residual issues; \textbf{M.V.} = Manually verified; 
\ding{108} = full agreement; $\LEFTcircle$ = partial agreement; 
``--'' = no finding; $\checkmark$ = any warning or error reported by any tool; \textbf{Y} = Yes.
}
\end{table*}

\label{sec:results-rq3}
\vspace{-0.2cm}
\begin{tcolorbox}[
    colframe=black,
    colback=gray!30,    
    boxsep=2pt,     
    left=4pt,
    right=4pt,
    top=2pt,
    bottom=2pt,
    boxrule=0.5pt,
]
\textbf{Answer to RQ3:} High-RRS benign functions are more frequently flagged by static analyzers, suggesting that residual similarity is a practical signal for prioritizing potential residual security-related issues.
\end{tcolorbox}

\vspace{-0.5cm}
\section{Empirical Validation of Residual Risk}
\label{sec:empirical}
\vspace{-0.3cm}
To complement the quantitative findings, we conduct an empirical validation to examine how residual risk manifests in practice. Since functions in the PrimeVul dataset are often isolated and lack necessary dependencies, static analysis tools cannot process them directly. To address this, we introduce minimal dummy helpers (e.g., type declarations, macros, and stub definitions) to reconstruct a compilable context without modifying the logic of the analyzed function (Fig. \ref{fig:dummy-helpers} in Appendix A). While these helpers provide the dependencies required for tool execution, they do not fully reproduce the original program context. Using this setup, we analyze high-RRS functions and identify \textbf{13} \textit{distinct categories} of errors and warnings, including memory safety violations, boundary issues, and null pointer dereferences (Table \ref{tab:validation_results}). The most frequent issues include \texttt{null pointer dereferences}, \texttt{uninitialized variable usage}, and \texttt{memory-related errors}. Table \ref{tab:risk_distribution} in Appendix A highlights the top 5 categories, emphasizing areas that require immediate attention. These results indicate that high residual similarity corresponds to code regions that retain security-relevant patterns even after patching.
We observe that many patches introduce minimal fixes (e.g., guard conditions) while leaving broader logic unchanged, leading static analyzers to flag potential issues such as null dereferences or buffer errors. This supports our hypothesis that RRS captures meaningful security-relevant properties. While not all flagged cases are exploitable and static analysis may produce false positives, they identify regions warranting further inspection. To improve robustness and reduce tool-specific bias, we employ three complementary state-of-the-art static analysis tools currently used in industry practice \cite{harmim2019scalable,thurnheer2021c++}. Our goal is not definitive vulnerability detection, but prioritization of high-risk candidates. Table~\ref{tab:validation_results} shows that multiple tools consistently identify residual issues across benign code, reinforcing the effectiveness of RRS-based prioritization. Finally, comparison with prior work in Table \ref{tab:paper_comparison}, illustrates that existing approaches focus primarily on pre-patch vulnerability detection or general similarity analysis, and do not explicitly address post-patch residual risk. This highlights the relevance of our empirical evaluation in targeting residual risk in benign code.

\vspace{-0.5cm}
\begin{table*}[htbp]
\centering
\scriptsize
\caption{Comparison with recent representative works on vulnerability analysis and code similarity.}
\vspace{0.2cm}
\label{tab:paper_comparison}
\begin{tabular}{l|>{\centering\arraybackslash}p{1.1cm}|>{\centering\arraybackslash}p{1.6cm}|>{\centering\arraybackslash}p{2.3cm}|>{\centering\arraybackslash}p{2.2cm}|>{\centering\arraybackslash}p{1.6cm}|>{\centering\arraybackslash}p{1.6cm}}
\hline

\textbf{Works}
& \textbf{Stage} 
& \textbf{Task} 
& \textbf{Technique} 
& \textbf{Representation Signals} 
& \textbf{Output} 
& \textbf{Post-Patch Risk}\\
\hline

\cite{cheng2022path}
& Pre-patch  
& VD 
& Graph learning (PDG) 
& Structural + PDG-based embeddings)
& Vulnerable / benign 
& $\times$ \\

\rowcolor{gray!30}
\cite{ebrahim2023} 
& Pre-patch 
& Code plagiarism 
& CodePTMs + classifiers 
& Semantic (embeddings) 
& Binary label 
& $\times$ \\

\cite{katz2025}
& Pre-patch  
& VD
& Embeddings + static analysis 
& Semantic + flow analysis 
& Vulnerability prediction 
& $\times$ \\

\rowcolor{gray!30}
\cite{martinez2025} 
& Pre-patch  
& Clone detection 
& Pretrained code models 
& Semantic 
& Similar / not similar 
& $\times$ \\

\cite{song2024revisiting} 
& General 
& Code similarity 
& AST edit distance 
& Structural (AST) 
& Similarity score 
& $\times$ \\

\rowcolor{gray!30}
\cite{tang2023}
& Post-patch  
& Patch classification 
& Multi-level embeddings 
& Semantic (multi-granular) 
& Patch label 
& $\times$ \\

\textbf{RRS}
& \textbf{Post-patch } 
& \textbf{Residual risk assessment}
& \textbf{Embeddings + AST + Model agreement}
& \textbf{Semantic + Structural + Consistency}
& \textbf{Risk score} 
& \textbf{\checkmark} \\

\hline
\end{tabular}
\vspace{0.3em}
\scriptsize{\\
\textbf{VD} = refers to vulnerability detection; $\checkmark$ = post-patch risk considered; $\times$ = not considered.
}
\end{table*}


\section{Threats to Validity}
\label{sec:threats}
\vspace{-0.3cm}

\noindent \textbf{Internal Validity.} Our validation relies on static analysis tools (Cppcheck, Clang-Tidy, Infer), which may produce false positives and negatives and do not capture runtime behavior. Although we mitigate this using multiple tools and manual checks, discrepancies may still affect reliability, and some flagged issues may reflect conservative heuristics rather than true vulnerabilities. \textbf{Construct Validity.} RRS is based on semantic similarity, structural similarity, and cross-model agreement, which may not capture all aspects of security risk. Embedding models capture semantic relationships but may not fully reflect all security-relevant behaviors, while AST-based similarity focuses on structural changes and may be influenced by parser-specific representations. \textbf{External Validity.} Our evaluation uses PrimeVul, which provides paired vulnerable--benign functions derived from real-world C/C++ vulnerabilities and is therefore well suited to this study. However, the dataset may not fully represent other programming languages, software ecosystems, or patching practices, potentially limiting generalizability.
\textbf{Selection Bias.} While residual issues are reported across all quadrants, our analysis emphasizes high-risk (Quadrant I) cases, which may underrepresent broader risk patterns in lower-risk regions. \textbf{Scalability.} To manage computational constraints, we limit AST analysis to bounded node sizes, which may exclude larger or more complex functions and affect completeness. \textbf{Interpretation of Residual Risk.} RRS provides a quantitative measure of residual proximity for prioritizing post-patch security review. High similarity often reflects localized patches that preserve intended functionality while flagging code regions for further inspection. Conversely, low similarity generally indicates greater code transformation, although additional analysis may still be needed to identify newly introduced defects or security weaknesses.

\vspace{-0.5cm}
\section{Related Work}
\label{sec:related_work}
\vspace{-0.2cm}
\noindent \textbf{Embedding-Based Code Similarity and VD.} Embedding-based similarity has been widely used in code analysis tasks such as plagiarism detection, clone detection, and vulnerability analysis \cite{ebrahim2023,tang2023}. Pretrained models (e.g., CodeBERT, UniXCoder) show strong performance but struggle to capture semantically equivalent code with structural differences \cite{martinez2025}. Hybrid approaches combining embeddings with static analysis further improve detection accuracy \cite{katz2025,farhad2025}. However, recent studies highlight opportunities to enhance embedding-based similarity by incorporating complementary signals \cite{nikiema2025small}.\\
\noindent \textbf{Structural and AST-Based Code Analysis.}
Structural representations (e.g., ASTs and program dependence graphs) complement embedding-based approaches, with graph-based methods like ContraFlow improving vulnerability detection and bug localization \cite{cheng2022path}, while AST-based representations are particularly effective for structurally sensitive tasks despite token-based models often performing better overall \cite{sun2023abstract}. Tree-based similarity metrics, such as AST edit distance, further demonstrate effectiveness in capturing structural similarity across languages \cite{song2024revisiting}. Additionally, neural approaches like UAST and CoCoAST enhance structural modeling for cross-language analysis and large-scale code tasks \cite{wang2022unified,shi2023cocoast}.\\
\noindent \textbf{Deep Learning for Software Testing and VD.}
Deep learning and LLMs are widely used for vulnerability detection, repair, and testing, though recent studies show their performance can be comparable to simpler models, suggesting reliance on shallow patterns \cite{weissberg2025llm}. They are also applied to automated repair (e.g., LoopRepair) and vulnerability assessment, where traditional ML and multi-task learning remain competitive and efficient \cite{ye2025well,nguyen2024automated}. Large-scale datasets support these approaches but often introduce noise, affecting reliability \cite{le2024automatic}, while recent work explores LLM-generated vulnerable code for data augmentation \cite{yi2026exploring}. Efforts to estimate residual risk further highlight the importance of prioritizing remaining vulnerabilities in software testing \cite{lee2026dependency}.\\
\noindent \textbf{Static Analysis and Post-Patch Security Assessment.}
Static analysis and empirical studies remain essential for vulnerability detection and post-patch evaluation, though challenges such as false positives and trust persist \cite{kholoosi2025software}. Comparisons show that static and learning-based approaches offer similar detection capabilities, with ML methods sometimes improving precision \cite{croft2021empirical}. Empirical evidence further reveals that vulnerabilities often persist in shared code and are not fully resolved through revisions \cite{selvaraj2022does}, while patch evolution over time complicates reliable post-patch security assessment \cite{xie2024unveiling}.

\vspace{-0.4cm}
\section{Conclusion \& Future Work}
\vspace{-0.2cm}
\label{sec:conc}

We revisit the assumption that patched code is inherently benign and introduce \texttt{Residual Risk Scoring (RRS)}, a framework that combines embedding-based similarity, localized AST-based structural analysis, and cross-model agreement. Our results show that benign functions often remain highly similar to their vulnerable counterparts, while static-analysis validation reveals that high-RRS functions frequently exhibit security-relevant issues. These findings suggest that representation-level similarity provides a practical signal for prioritizing post-patch inspection and security review. Future work will extend RRS to additional datasets, programming languages, dynamic analysis techniques and CI/CD-based post-patch review workflows. Overall, our findings highlight the value of representation-aware analysis for supporting software security assurance through vulnerability triage, patch validation, and risk-aware software maintenance.

\section{Acknowledgement}
This research was funded in part by U.S. National Science Foundation under Grant Number OIA-2437963 and Louisiana Board of Regents.

%
%
\bibliographystyle{splncs04}
\bibliography{bibliography}
\appendix

\section{Appendix}
\label{sec:appendix}
\vspace{-0.8cm}

\begin{figure}[htbp]
\centering

\begin{minipage}[t]{0.48\linewidth}
\centering
\vspace{0pt}
\includegraphics[width=\linewidth]{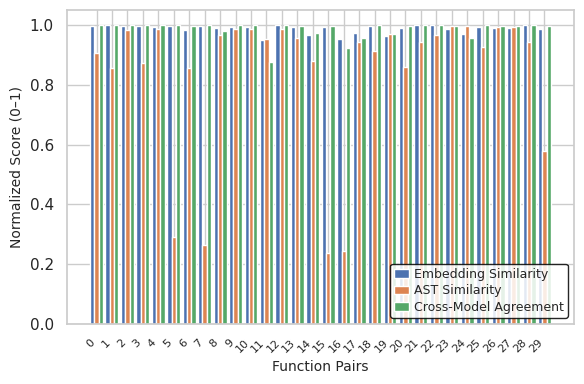}
\caption{Visualization of embedding similarity, localized AST similarity, and cross-model agreement across a subset of vulnerable–benign function pairs.}
\label{fig:multisignal-bars}
\end{minipage}
\hfill
\begin{minipage}[t]{0.48\linewidth}
\centering
\vspace{0pt}

\begin{subfigure}[t]{0.48\linewidth}
\centering
\includegraphics[width=\linewidth,height=3.4cm]{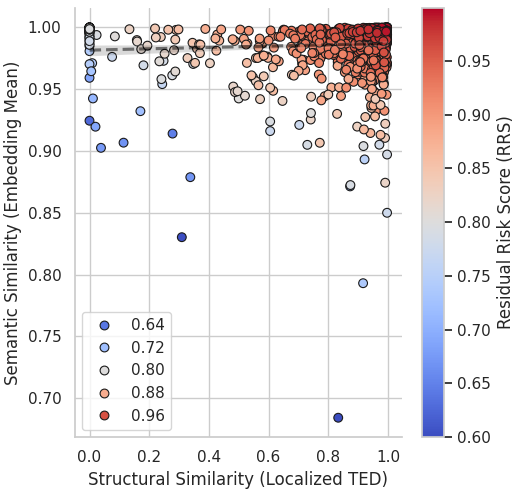}
\caption{$\alpha=0.6, \beta=0.2$}
\end{subfigure}
\hfill
\begin{subfigure}[t]{0.48\linewidth}
\centering
\includegraphics[width=\linewidth,height=3.4cm]{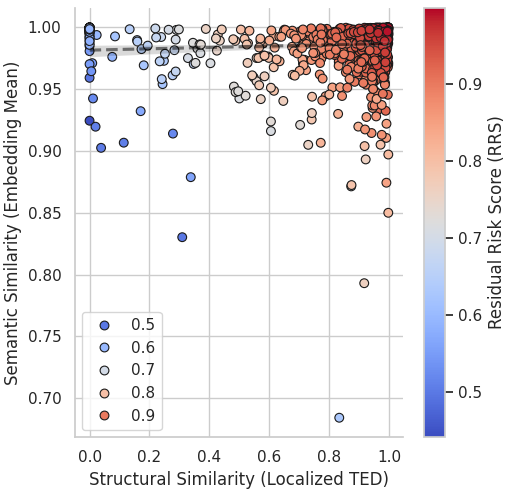}
\caption{$\alpha=0.2, \beta=0.6$}
\end{subfigure}

\caption{Sensitivity analysis under $(\alpha,\beta)$ configurations: $(0.6, 0.2)$ and $(0.2, 0.6)$.}
\label{fig:d-e}

\end{minipage}

\end{figure}
\vspace{-0.7cm}
\begin{figure}[htbp]
\centering
\begin{minipage}[t]{0.48\linewidth}
    \centering
    \vspace{0pt}
    
    \includegraphics[width=\linewidth]{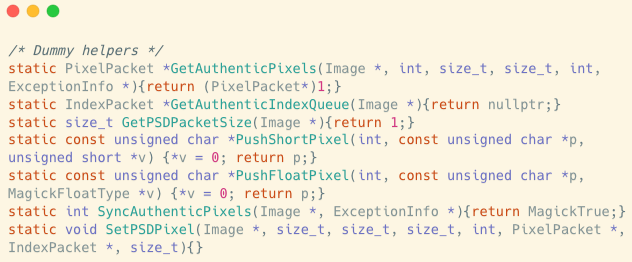}
    \caption{Dummy helpers used for static analysis.}
    \label{fig:dummy-helpers}
\end{minipage}
\hfill
\begin{minipage}[t]{0.48\linewidth}
    \centering
    \vspace{0pt}
    \scriptsize
    \captionof{table}{Top five distribution of residual risk issues identified by static analysis tools.}
    \label{tab:risk_distribution}
    \begin{tabular}{l|c|c}
    \hline
    \textbf{Residual Risk Type} & \textbf{Frequency} & \textbf{Priority} \\
    \hline
    Null pointer dereference & High & \redCircle \\
    Uninitialized variable & High & \redCircle \\
    Memory / resource leak & Medium & \lightredCircle \\
    Allocation overflow & Medium & \yellowCircle \\
    Control-flow issues & Low & \lightyellowCircle \\
    \hline
    \end{tabular}
\end{minipage}
\hfill
\begin{minipage}[t]{0.48\linewidth}
    \centering
    \vspace{0pt}
    \scriptsize
    \captionof{table}{Residual issues across RRS quadrants.}
    \label{tab:residual-issues-Q}
    \begin{tabular}{l|c}
    \hline
    \textbf{Quadrant} & \textbf{Residual Issues (\%)} \\
    \hline
    I ($E_H$, $A_H$)  & \cellcolor{gray!30} 61\% \\
    II ($E_H$, $A_L$)  & 11\% \\
    III ($E_L$, $A_H$)  & 9\% \\
    IV ($E_L$, $A_L$) & 6\% \\
    \hline
    \end{tabular}
\end{minipage}
\end{figure}

\end{document}